\documentclass[11pt,a4paper]{JHEP3}
\usepackage{amsmath, amssymb}
\usepackage{verbatim}
\usepackage{latexsym}
\usepackage{euscript}
\usepackage{amsfonts}
\usepackage{verbatim}
\def\be{\begin{eqnarray}}
\def\ee{\end{eqnarray}}
\def\0{\nonumber}

\newcommand\ER{\EuScript{R}}
\newcommand\ES{\EuScript{S}}
\newcommand\EE{\EuScript{E}}
\newcommand\ED{\EuScript{D}}
\usepackage[]{array}
\usepackage[]{amsmath}
\usepackage[]{amssymb}
\usepackage[]{mathrsfs}
\usepackage[]{tensor}
\usepackage[symbol]{footmisc}
\newcommand{\del}{\partial}

\newcommand{\al}{\alpha}
\newcommand{\dal}{\dot\alpha}
\newcommand{\Si}{\Sigma}
\newcommand{\bS}{\bar\Sigma}
\newcommand{\gam}{\gamma}
\newcommand{\dbeta}{\dot\beta}
\newcommand{\dgamma}{\dot\gamma}
\newcommand{\ddelta}{\dot\delta}
\newcommand{\deps}{\dot\epsilon}
\newcommand{\eps}{\epsilon}

\newcommand{\La}{\Lambda}
\newcommand{\bL}{\bar\Lambda}

\preprint{SISSA/20/2013/FISI\\\tt hep-th/1305.7116\\revised version}

\title{Weyl transformations and trace anomalies in N=1, D=4 supergravities}

\author{ L.Bonora\\
International School for Advanced Studies (SISSA)\\
Via Bonomea 265, 34136 Trieste, Italy, and INFN, Sezione di
Trieste, Italy;\\
E-mail: \email{bonora@sissa.it},}

\author{ S.Giaccari\\
International School for Advanced Studies (SISSA)\\
Via Bonomea 265, 34136 Trieste, Italy, and INFN, Sezione di
Trieste, Italy;\\
E-mail:   \email{giaccari@sissa.it},}

\abstract{
We consider the supersymmetric extension of Weyl transformations in various types of supergravities, the minimal, nonminimal and new minimal N=1 SUGRA in 4D, formulated in terms of superfields and study their local cohomology. Based also on previous results we conclude that there are only two types of trace anomalies in nonminimal and new minimal supergravities, which correspond to the two nontrivial cocycles of the minimal supergravity and, when reduced to component form, to the well-known squared Weyl density and Euler density.}


\begin{document}
\section{Introduction}

In this paper we want to analyze the supersymmetric extension of Weyl transformations in various types of supergravities, the minimal, non minimal and new minimal N=1 SUGRA in 4D, and study the general structure of trace anomalies. To this end, rather than considering specific cases we carry out a cohomological analysis, whose validity is not limited to one-loop calculations. 

The motivation for this research is twofold. On the one hand it has been pointed out recently that `old' minimal supergravity in 4D (with 12+12 dofs) might be inconsistent due to the presence of an inherent global conserved current, \cite{KS}. It has also been suggested that a different type of SUGRA, referred to henceforth as {\it new minimal}, characterized by 16+16 dofs, may be exempt from this risk. This model has been identified with the supergravities studied in \cite{GGMW2} and \cite{LLO}. The study of conformal anomalies in these and other models is interesting not only in itself, but also because it allows us to identify what the `superWeyl group' is, as will be seen below.

Another motivation arises from the proposal of \cite{Nakayama} that a source of CP violation in a 4D theory coupled to gravity could come from the trace anomaly. The trace anomaly may contain, in principle, beside the Weyl density (square of the Weyl tensor)
\be
\ER_{nmkl} \ER^{nmkl}-2 \ER_{nm}\ER^{nm} +\frac 13 \ER^2\label{weyl}
\ee
and the Gauss-Bonnet (or Euler) one
\be
\ER_{nmkl} \ER^{nmkl}-4 \ER_{nm}\ER^{nm} + \ER^2,\label{gausbonnet}
\ee
another nontrivial piece, the Pontryagin density
\be
\epsilon^{nmlk}\ER_{nmpq}\ER_{lk}{}^{pq}\label{Pontryagin}
\ee
Each of these terms appears in the trace of the e.m. tensor with its own coefficient. The first two are denoted $c$ and $a$, respectively. They are known at one-loop for any type of (Gaussian) matter \cite{DC, CD1}, and $a$ is the protagonist of recent important developments, \cite{KSch}. The coefficient of (\ref{Pontryagin}) is not sufficiently studied. 
The original purpose of this paper was to analyse whether the appearance of such a term in the trace anomaly is compatible with supersymmetry. Since it is hard to supersymmetrize these three pieces and relate them to one another in a supersymmetric context, the best course is to proceed in another way, that is to consider a conformal theory in 4D coupled to (external) supergravity formulated in terms of superfields and find all the potential superconformal anomalies. This will allow us to see whether (\ref{Pontryagin}) can be accommodated in an anomaly supermultiplet as a trace anomaly member. 

This type of analysis was carried out long ago for minimal supergravity, see \cite{BPT} and also \cite{Brandt2}. Our purpose here is to extend it to other types of 4D supergravities \footnote{A general cohomological treatment of anomalies in 4D supergravities, which has some overlap with the present paper, is contained  in \cite{Brandt1}.}, in particular to the new minimal SUGRA mentioned above, \cite{GGMW2} and \cite{LLO}. Unfortunately there is no unique choice of the torsion constraints for these theories and no unique superfield formalism, (see
\cite{StelleWest,FerVan} and \cite{AVS,SoniusWest} for earlier `minimal' formulations and 
\cite{FGKVP} for their equivalence; see \cite{Breit,SiegelGates} for earlier non-minimal formulations; see moreover \cite{Buchbinder,Butter1,Butter2} and the textbooks \cite{GGRS,Kuzenko}). Thus we have chosen to follow the formalism of \cite{WB}, further expanded in \cite{GGMW1,GMW,GGMW2}. Our original aim, the analysis of trace anomalies, has turned out to be anything but standard, contrary to the case of minimal supergravity. The reason is that in the latter case the cohomological analysis can be done on a differential space formed by polynomials of the superfields. In the other abovementioned versions of supergravities one has to admit in the differential space also nonpolynomial expressions of the superfields, due to the essential role of dimensionless prepotentials in these models. To solve in a satisfactory way the cohomology problem one has to start from minimal supergravity and map its cocycles to the other models with the superfield mappings of ref.\cite{GGMW1,Kuzenko}. Once this is clarified the possible superconformal anomalies are rather easily identified. Based also on the analysis carried out long ago in \cite{BPT}, one can conclude that there are, not unexpectedly, two independent anomalies corresponding to the square Weyl and Gauss-Bonnet densities, much like in minimal supergravity. The anomaly corresponding to the Gauss-Bonnet density has a particularly complicated form in non-minimal and new minimal supergravities, and could be identified only via the abovementioned mapping method.

The conclusion concerning the Pontryagin density (\ref{Pontryagin}) is negative: in all types of supergravities the Pontryagin 
density does not show up in the trace anomaly, but it appears in the chiral (Delbourgo-Salam) anomaly, which, as expected, belongs, together with the trace anomaly, to a unique supermultiplet. 

The paper is organized as follows. In the next section we review the minimal supergravity case. This section does not contain new results, it has mainly pedagogical and reference purposes: the analysis is far clearer if we keep in mind the minimal case as a guide. In section 3 we introduce the non minimal supergravity and the relevant superconformal transformations. We also easily identify a non-trivial cocycle corresponding to the square Weyl density. In section 4 we do the same for the new minimal supergravity. In section 5 we study the reduction to component form of the latter cocycle. In section 6 we briefly introduce the mapping method from one supergravity model to another and introduce the relevant formulas. In section 7 we move on to compute the remaining superWeyl cocycles. Section 8 is devoted to the conclusions.

\section{N=1 minimal supergravity in D=4 and its superfields}

For the notation we follow \cite{WB}. The superspace of $N=1$ supergravity is spanned by the supercoordinates $Z^M=(x^m,\theta^\mu,\bar\theta_{\dot\mu})$.
The minimal $N=1$ supergravity in $D=4$ can be formulated in terms of the superfields: $R(z),G_a(z)$ and $W_{\alpha\beta\gamma}(z)$. $R$ and $W_{\alpha\beta\gamma}$ are chiral while $G_a$ is real. We will also need the antichiral superfields $R^+(z)$ and $\bar W_{\dot\alpha\dot\beta\dot\gamma}(z)$, conjugate to $R$ and $W_{\alpha\beta\gamma}$, respectively. $W_{\alpha\beta\gamma}$ is completely symmetric in the spinor indices $\alpha,\beta,...$.  These superfields are subject to the constraints:
\be 
&& \nabla^{\al} G_{\al \dot\beta}=\bar \nabla_{\dbeta} R^+,\quad\quad\bar\nabla^{\dot\beta}G_{\alpha\dot\beta}=\nabla_\alpha R\0\\
&&\nabla^\al W_{\al\beta\gamma}+\frac i2 ( \nabla_{\beta\dbeta} G_{\delta}{}^{\dbeta}+ \nabla_\beta\ddelta G_{\dbeta}{}^{\dbeta})=0\0\\
&&\bar\nabla^{\dal} W_{\dal\dbeta\dgamma}+\frac i2 ( \nabla_{\beta\dbeta} G^\beta{}_{\ddelta}+ \nabla_\beta{}^{\ddelta} G^\beta{}_{\dbeta})=0\label{constraints}
\ee
The latter are found by solving the (super)Bianchi identities for the supertorsion and
the supercurvature
\be
&&T^A=dE^A+E^B\phi_B{}^A=\frac 12 E^CE^BT_{BC}{}^A= \frac 12 dz^Mdz^NT_{NM}{}^A\label{torsion}\\
&& R_A{}^B=\frac 12 E^CE^DR_{DCA}{}^B=dz^Mdz^N \partial_N \phi_{MA}{}^B+ dz^M\phi_{MA}{}^C dz^N \phi_{NC}{}^B\label{curvature}\0
\ee
where $\phi_{MA}{}^B$ is the superconnection and $E^A=dz^M E_M{}^A$ the supervierbein
\be
E_M{}^AE_A{}^N= \delta_M{}^N,\quad\quad E_A{}^ME_M{}^B=\delta_A{}^B,\0
\ee
after imposing by hand the restrictions
\be
&&T_{\underline\al \underline \beta}{}^{\underline \gamma}=0,\quad T_{\al\beta}{}^c=T_{\dal\dbeta}{}^c=0
\0\\
 &&T_{\al\dbeta}{}^c=T_{\dbeta\al}{}^c= 2i\sigma_{ \al\dbeta}{}^c\0\\
&&T_{\underline\al b}{}^c=T_{b \underline\al }{}^c=0, \quad\quad T_{ab}{}^c=0\label{constr}
\ee
where $\underline {\alpha}$ denotes both $\al$ and $\dal$.
The superdeterminant of the vierbein $E_M{}^A$ will be denoted by $E$.

The Bianchi identities are
\be
{\cal D}{\cal D} E^A =E^BR_B{}^A, \quad\quad \nabla T^A=E^BR_B{}^A\label{bianchi}
\ee
where ${\cal D}= dz^M \nabla_M$ and $\nabla_A=E_A{}^M \nabla_M$.
Imposing (\ref{bianchi}) one gets all the components of $T^A$ and $R_A{}^B$ in terms of $R,G_a, W^{\al\beta\gamma}$ and their conjugates. The other Bianchi identity
\be
({\cal D} R)_A{}^B=0\label{bianchi2}
\ee
is automatically satisfied.

\subsection{Superconformal symmetry and (super)anomalies}
\label{app:superconf}

Superconformal transformations are defined by means of the chiral superfield parameter $\sigma=\sigma(z)$ and its conjugate $\bar \sigma$.
\be
&&\delta E_M{}^a= (\sigma +\bar \sigma)  E_M{}^a\label{superconf}\\
&&\delta E_M{}^{\al}= (2\bar\sigma - \sigma)  E_M{}^{\al}+\frac i2 E_M{}^a \bar \sigma_a^{\dal\al}\nabla_{\dal}\bar \sigma\0\\
&&\delta E_M{}^{\dal}= (2\sigma - \bar\sigma)  E_M{}^{\dal}+\frac i2 E_M{}^a \bar \sigma_a^{\dal\al}\nabla_{\al} \sigma\0\\
&&\delta \phi_{M\al\beta} = E_{M\al} \nabla_\beta \sigma+E_{M\beta} \,\nabla_\alpha \sigma+
(\sigma^{ab})_{\al\beta}E_{Ma}\, \nabla_b(\sigma+\bar\sigma)\0
\ee
where
\be
 \phi_{M\al}{}^{\beta}= \frac 12 \phi_{Mab} (\sigma^{ab})_\al{}^\beta,
\quad\quad \phi_{M}{}^{\dal}_{\dbeta}= \frac 12 \phi_{Mab} (\bar\sigma^{ab})^{\dal}{}\beta\0
\ee
The transformations (\ref{superconf}) entail
\be
&&\delta E =2(\sigma+\bar\sigma) E\label{superfieldconf}\\
&& \delta R =(2\bar\sigma-4\sigma)R-\frac 14 \nabla_{\dal}\nabla^{\dal} \bar\sigma\0\\
&& \delta R^+ =(2\sigma-4\bar\sigma)R^+-\frac 14 \nabla^{\al}\nabla_{\al}\sigma\0\\
&&\delta G_a= -(\sigma+\bar \sigma)G_a +i \nabla_a(\bar \sigma-\sigma)\0\\
&&\delta W_{\al\beta\gamma} = -3\sigma W_{\al\beta\gamma}\0
\ee
If we promote the superfield $\sigma$ to a superghost superfield, by inverting the spin-statistics connection, so that it becomes an anticommuting parameter, it is easy to prove that the above transformations are nilpotent.

Let us define the functional operator that implements these transformations, i.e.
\be
\Sigma =\int_{x\theta} \delta\chi_i\, \frac {\delta}{\delta\chi_i}\0
\ee
where $\chi_i$ represent the various superfields in the game and $_{x\theta}$ denotes integration 
$d^4x d^4\theta$. This operator is nilpotent: $\Sigma^2=0$.
As a consequence it defines a cohomology problem. The cochains are integrated local expressions of the superfields and their superderivatives, invariant under superdiffeomorphism and local superLorentz transformations. Candidates for superconformal anomalies are nontrivial cocycles of $\Sigma$ which are not
coboundaries, i.e. integrated local functionals $\Delta_\sigma$, linear in $\sigma$, such that
\be
\Sigma\, \Delta_\sigma=0, \quad \quad {\rm and}\quad\quad \Delta_\sigma\neq \Sigma\, {\cal C}\label{anom}
\ee
for any integrated local functional ${\cal C}$ (not containing $\sigma$).

The complete analysis of all the possible nontrivial cocycles of the operator $\Sigma$ was carried out long ago in \cite{BPT}. It was shown there that the latter can be cast into the form
\be
\Delta_\sigma= \int_{x\theta} \left[ \frac {E(z)}{-8R(z)}\,\sigma(z)\, \ES(z) + h.c.\right]\label{Deltasigma}
\ee
where $\ES(z)$ is a suitable chiral superfield. In \cite{BPT} all the possibilities for $\ES$ were classified.
For pure supergravity (without matter) the only nontrivial possibilities turn out to be:
\be
\ES_1(z)= W^{\al\beta\gamma}W_{\al\beta\gamma} \quad\quad {\rm and} \quad\quad 
\ES_2(z) = (\bar\nabla_{\dal}\bar \nabla^{\dal} -8R) (G_aG^a+2RR^+)\label{S1S2}
\ee
(the operator $(\bar\nabla_{\dal}\bar \nabla^{\dal} -8R)$ maps a real superfield into a chiral one).

It is well-known that the (\ref{Deltasigma}) cocycles contain not only the trace anomaly, but a full supermultiplet of anomalies. The local expressions of the latter are obtained by stripping off the corresponding parameters from the integrals in  (\ref{Deltasigma}). Let us recall also 
that the conversion of $\sigma$ to an anticommuting parameter is not strictly necessary: eq.(\ref{anom}) simply corresponds to the Wess-Zumino consistency conditions, i.e. to the invariance under reversing the order of two successive (Abelian) gauge transformations. But an anticommuting $\sigma$ allows us to use the incomparably simpler formalism of cohomology.

\subsection{Meaning of superconformal transformations}
\label{app:meaning}

Eqs. (\ref{S1S2}) are rather implicit and it is opportune to see the corresponding expressions in component fields, at least as far as the dependence on the metric alone is concerned. This reduction has been done in \cite{AFGJ}. We repeat it here for pedagogical reasons, but also because the formalism we use is different from the one of \cite{AFGJ}. The method below will be used throughout the paper. In general the expressions of the above cocycles in components
are extremely complicated and really unmanageable because of the presence of auxiliary fields. We are interested in recognizing the two cocycles (\ref{S1S2}) when only the metric is taken into account while all the other fields are ignored, so that we can compare them with the usual Weyl cocycles (the squared Weyl tensor, the Gauss-Bonnet and the Pontryagin densities). Our task in the sequel is to extract such expressions from (\ref{S1S2}). We will refer to them as the {\it ordinary} parts of the cocycles.

We first introduce the relevant components fields and clarify the meaning of 
the components in the parameters superfield $\sigma(z)$. To start with let us define the lowest component fields of the supervierbein as in \cite{WB}
\be\label{vier1}
E_M{}^A(z){\big\vert}_{\theta=\bar\theta=0} = \left(\begin{matrix} 
e_m{}^a(x) & \frac 12 \psi_m{}^{\al}(x)& \frac 12 \bar \psi_{m\dal}(x)\\
0& \delta_\mu{}^\alpha & 0\\
0&0& \delta^{\dot\mu} {}_{\dal}
	                                             \end{matrix}\right)
\ee
and
\be\label{vier2}
E_A{}^M(z){\big\vert}_{\theta=\bar\theta=0} = \left(\begin{matrix} 
e_a{}^m(x) & -\frac 12 \psi_a{}^{\mu}(x)& -\frac 12 \bar \psi_{a\dot\mu}(x)\\
0& \delta_\al{}^\mu & 0\\
0&0& \delta^{\dal} {}_{\dot\mu}
	                                             \end{matrix}\right)
\ee
where $e_m{}^a$ are the usual 4D vierbein and $\psi_m{}^{\al}(x),\bar \psi_{m\dal}(x)$ the gravitino field components. We have in addition
\be 
R(z){\big\vert}_{\theta=\bar\theta=0}= -\frac 16 M(x), \quad\quad G_a(z) {\big\vert}_{\theta=\bar\theta=0}= -\frac 13 b_a(x)\label{aux}
\ee
where $M$ is a complex scalar field and $b_a$ is a real vector field.
As for the superconnection we have
\be\label{phi}
\phi_{mA}{}^B{\big\vert}_{\theta=\bar\theta=0}= \omega_{mA}{}^B(x),\quad\quad
\phi_{\mu A}{}^B{\big\vert}_{\theta=\bar\theta=0}= 0,\quad\quad \phi_{\dot\mu A}{}^B{\big\vert}_{\theta=\bar\theta=0}= 0,
\ee
and $\omega_{mA}{}^B(x)$ is of course of the Lorentz type. Its independent components turn out to be
\be
&&\omega_{nml}\equiv e_m{}^a e_{lb} \omega_{na}{}^b=\label{omega}\\
&& = \frac 12 \Bigl[e_{na}(\del_me_l{}^a-\del_l e_m{}^a)-e_{la}(\del_n e_m{}^a-\del_m e_n{}^a)
-e_{ma}(\del_l e_n{}^a-\del_n e_l{}^a)\Bigr]\0\\
&&+ \frac i4 \Bigl[e_{na} \left(\psi_l \sigma^a \bar\psi_m-\psi_m \sigma^a \bar\psi_l\right)
- e_{la} \left(\psi_m \sigma^a \bar\psi_n-\psi_n \sigma^a \bar\psi_m\right)
- e_{ma} \left(\psi_n \sigma^a \bar\psi_l-\psi_l \sigma^a \bar\psi_n\right)\Bigr]\0
\ee
This has the same symmetry properties in the indices as the usual spin connection and reduces to it when the gravitino field is set to 0.

It is then easy to prove, using (\ref{curvature}), that 
\be
R_{nma}{}^b{\big\vert}_{\theta=\bar\theta=0}= \del_n \omega_{ma}{}^b-\del_m \omega_{na}{}^b+
\omega_{ma}{}^c\omega_{nc}{}^b-\omega_{na}{}^c\omega_{mc}{}^b\equiv \ER_{nma}{}^b\label{Riemann}
\ee
This relation will be used later on. In conclusion the independent component fields are the vierbein, the gravitino and the two auxiliary fields $M$ and $b_a$.

Let us come now to the interpretation of the superconformal transformations (\ref{superconf}).
To this end we expand the chiral superfield $\sigma(z)$ in the following way:
\be 
\sigma(z)= \omega(x)+i \alpha(x) + \sqrt 2 \Theta^{\al} \chi_\alpha(x) + \Theta^{\al}\Theta_\alpha (F(x)+iG(x))\label{sigma}
\ee
where we have introduced new anticommuting variables $\Theta^\alpha$, which, unlike $\theta^\mu$, carry Lorentz indices. This is always possible, see \cite{WB}: the first term on the RHS corresponds
to $\sigma{\big\vert}_{\theta=\bar\theta=0}$, $\chi_\alpha$ to $\nabla_\alpha \sigma {\big\vert}_{\theta=\bar\theta=0}$, and $F(x)+iG(x)$ to $\nabla^\alpha\nabla_\alpha \sigma{\big\vert}_{\theta=\bar\theta=0}$.
Comparing now with the first equation in (\ref{superconf}), and taking into account (\ref{vier1},\ref{vier2}), we see that $\omega(x)$ is the parameter of the ordinary Weyl transformation, while comparing with the second and third equation in  (\ref{superconf}) one can see that $\psi_\alpha$
and $\bar \psi^{\dal}$ transform with opposite signs with respect to the parameter $\alpha(x)$. Thus
$\alpha(x)$ is the parameter of an ordinary chiral transformation.

Therefore when (\ref{S1S2}) is inserted in (\ref{Deltasigma}) the term linear in $\omega(x)$ will represent a conformal anomaly, while the term
linear in $\alpha(x)$ will represent a chiral (Delbourgo-Salam) anomaly. Similarly the term linear
in $\chi_\alpha$ is the supercurrent anomaly. For the meaning of the cocycles linear in $F(x)$ and $G(x)$ see for instance \cite{AFGJ}. Not surprisingly all these anomalies form an $N=1$ supermultiplet.

The next step is to derive the conformal and chiral anomalies in components.

\subsection{Anomalies in components}
\label{app:companom}

To derive the anomalies in components we have to integrate out the anticommuting variables. To this end it is convenient to use, instead of the superdeterminant $E$, the chiral density $\EE$ (see \cite{WB}). The latter is defined by
\be 
\EE(z)= a(x) + \sqrt 2 \Theta  \rho(x) + \Theta\Theta f(x)\label{chiraldensity}
\ee
where $a(x)= \frac 12 e(x) \equiv \frac 12 \det e_m{}^a$. The $\rho$ and $f$ components contain, beside $e$
the gravitino and/or the auxiliary field $M$, and they vanish when the latter are set to 0.
We can rewrite our two integrated cocycles as follows
\be
\Delta_\sigma^{(i)}= \int d^4x \left( \int d^2\Theta\, \EE(z) \,\sigma(z)\, \ES_i(z)\,+\,h.c.\right), \quad\quad
i=1,2\label{Deltasigmai}
\ee
This means that, given the interpretation of the lowest components of $\sigma(z)$ as the parameters
of the conformal and chiral transformations, and due to (\ref{chiraldensity}), the {\it ordinary} part of the conformal
and chiral anomaly terms (i.e. the terms linear in $\omega$ and $\alpha$) will depend on $\nabla\nabla \ES_i\equiv \nabla^\alpha\nabla_\alpha \ES_i$, because this corresponds to the coefficient of $\Theta\Theta$ in the expansion of $\ES_i$. 
So finally we can write
\be
\Delta_\sigma^{(i)}\approx 4\int d^4x \left(\frac 12 e \, (\omega +i \alpha)\,\nabla\nabla S_i{\big\vert}_{\theta=\bar\theta=0}\,+\,h.c.\right), \quad\quad
i=1,2\label{Deltasigmax}
\ee
where $\approx$ means `up to terms that vanish when all the fields except the metric are set to 0'. The anomalous trace of the energy-momentum tensor and the divergence of the chiral current are obtained from the integral on the RHS of (\ref{Deltasigmax}) by stripping off it the parameters $\omega$ and $\alpha$, respectively.

\subsubsection{The square Weyl cocycle}

Let us start from $S_1$. The relevant terms to be considered are $\nabla^\alpha \nabla_\alpha W_{\beta\gamma\delta} W^{\beta\gamma\delta}$ and $ \nabla^\alpha W^{\beta\gamma\delta}
\nabla_\alpha W_{\beta\gamma\delta}$ at $\theta=\bar\theta=0$. The term $W_{\beta\gamma\delta}{\big\vert}_{\theta=\bar\theta=0}$ is linear in the gravitino field and
in the field $b_a$. As a consequence this term does not affect the ordinary part of the anomaly.
On the contrary the square derivative of $W$ does affect the ordinary part of the anomaly. It is therefore
necessary to compute it explicitly. The symmetric part of $\nabla_\alpha W_{\beta\gamma\delta}{\big\vert}_{\theta=\bar\theta=0}$ can be computed as follows. Let us consider 
the identity
\be
R_{nma}{}^b= E_n{}^cE_m{}^d R_{cda}{}^b +E_n{}^{\underline \gamma}E_m{}^d R_{\underline \gamma da}{}^b+ E_n{}^cE_m{}^{\underline\delta} R_{c\underline \delta a}{}^b-E_n{}^{\underline\gamma}E_m{}^{\underline \delta} R_{{\underline\gamma}{\underline \delta}a}{}^b\label{Rscomposto}
\ee
and evaluate it at $\theta=\bar\theta=0$. We know the LHS due to (\ref{Riemann}). The RHS contains various expressions, and in particular the totally symmetrized derivative $\nabla_{(\alpha} W_{\beta\gamma\delta)}$.
It is possible to project it out and get
\be
\nabla_{(\alpha} W_{\beta\gamma\delta)}{\big\vert}_{\theta=\bar\theta=0}= -\frac 1{16} (\sigma^a\bar\sigma^b\epsilon)_{(\al\beta} 
 (\sigma^c\bar\sigma^d\epsilon)_{\gamma\delta)}\ER_{abcd}\label{dWsymm}
\ee 
and similarly
\be
\nabla_{(\dal} W_{\dbeta\dgamma\ddelta)}{\big\vert}_{\theta=\bar\theta=0}= -\frac 1{16} (\epsilon\bar\sigma^a\sigma^b)_{(\dal\dbeta} 
 (\epsilon\bar\sigma^c\sigma^d)_{\dgamma\ddelta)}\ER_{abcd}\label{dbarWsymm}
\ee 
where $\ER_{abcd}= e_a{}^ne_b{}^m \ER_{nmcd}$.

Using the second equation in (\ref{constraints}) one can easily obtain
\be 
\nabla_{\al} W_{\beta\gamma\delta}&=&\nabla_{(\al} W_{\beta\gamma\delta)} \label{symmetrized}\\
&&+\frac i4 \left(
\epsilon_{\al\beta}(\sigma^{ab}\epsilon)_{\gamma\delta}+ \epsilon_{\al\gamma}(\sigma^{ab}\epsilon)_{\beta\delta}+
\epsilon_{\al\delta}(\sigma^{ab}\epsilon)_{\beta\gamma}\right)(\nabla_a G_b-\nabla_b G_a)\0
\ee
and a similar equation for the conjugate derivative. Now let us see, as an example of arguments that will be repeatedly used in the sequel, that $\nabla_a G_b-\nabla_b G_a$ evaluated at $\theta=\bar\theta=0$ does not contribute to the ordinary part of the anomaly. In fact
 $\nabla_a G_b$ cannot contribute to it, for we have
\be
\nabla_a G_b= E_a{}^M \del_M G_b +  E_a{}^M \phi_{Mb}{}^c G_c\0
\ee
The last term, when evaluated at $\theta=\bar\theta=0$ is linear in the field $b_c$. The second
term in the RHS can be written
\be
E_a{}^M \del_M G_b{\big\vert}_{\theta=\bar\theta=0}=-\frac 13 e_a{}^m \del_mb_a  -\frac 12 e_a{}^m \psi_m{}^\alpha
\nabla_\alpha G_b{\big\vert} -\frac 12 e_a{}^m \bar\psi_{m\dal}\nabla^{\dal} G_b{\big\vert}\0
\ee
where the vertical bar stands for ${\big\vert}_{\theta=\bar\theta=0}$.
Since both $\nabla_\alpha G_b{\big\vert}$ and $\nabla^{\dal} G_b{\big\vert}$ vanish when the gravitino and the auxiliary fields are set to 0, it follows that also $\nabla_a G_b$ vanishes in the same circumstances.
Therefore for our purposes only the completely symmetrized spinor derivative of $W$ matters in (\ref{symmetrized}). We will write
\be\label{approx}
\nabla_{\al} W_{\beta\gamma\delta}\approx\nabla_{(\al} W_{\beta\gamma\delta)}, \quad\quad
\nabla_{\dal} W_{\dbeta\dgamma\ddelta}\approx\nabla_{(\dal} W_{\dbeta\dgamma\ddelta)}
\ee
to signify that the LHS is equal to the RHS up to terms that vanish when the gravitino and the auxiliary fields are set to 0.

Now it is a lengthy but standard exercise to verify that
\be
\nabla^\alpha W^{\beta\gamma\delta}
\nabla_\alpha W_{\beta\gamma\delta}{\big\vert} \approx \frac 18 \Bigl( \ER_{nmkl} \ER^{nmkl}-2 \ER_{nm}\ER^{nm} +\frac 13 \ER^2+
\frac i2 \epsilon^{nmlk}\ER_{nmcd}\ER_{lk}{}^{cd}\Bigr)\label{DWDW}
\ee
where $\ER_{nmkl}=e_n{}^ae_m{}^be_k{}^ce_l{}^d \ER_{abcd}$, $\ER_{nm}=e_k{}^a e^{kb}\ER_{anbm}$ and
$\ER= e_n{}^a e^{nc} e_m{}^b e^{md} \ER_{abcd}$. The first three terms in brackets in the RHS are easily recognized
to correspond to the ordinary Weyl density, while the fourth term is the Pontryagin density. We thus have
\be
\Delta_\sigma^{(1)}&\approx& 4\int d^4x \, e \Bigl[(\omega + i \alpha)\, \nabla^\alpha W^{\beta\gamma\delta}
\,\nabla_\alpha W_{\beta\gamma\delta}{\big\vert}+h.c.\Bigr]\0\\
&\approx&\frac 1{2} \int d^4x \, e \Bigl[(\omega + i \alpha) \Bigl( \ER_{nmkl} \ER^{nmkl}-2 \ER_{nm}\ER^{nm} +\frac 13 \ER^2+\frac i2 \epsilon^{nmlk}\ER_{nmcd}\ER_{lk}{}^{cd}\Bigr)+ h.c.\Bigr]\0\\
&=&\int d^4x \, e \Bigl\{\omega\biggl( \ER_{nmkl} \ER^{nmkl}-2 \ER_{nm}\ER^{nm} +\frac 13 \ER^2\biggr) - \frac 12\, \alpha \, \epsilon^{nmlk}\ER_{nmpq}\ER_{lk}{}^{pq}\Bigr\}\label{Weyl+SS}
\ee
In the last line one recognizes the conformal Weyl anomaly linear in $\omega$ and the Delbourgo-Salam anomaly linear in $\alpha$.

\subsubsection{The Gauss-Bonnet cocycle}

The second cocycle is determined by $\nabla\nabla \ES_2{\big\vert}_{\theta=\bar\theta=0}$ and its hermitean conjugate. 
Since $G_a, R,R^+$ and their first order spinorial derivative evaluated at $\theta=\bar\theta=0$ all vanish when
the gravitino and auxiliary fields are set to 0, the {\it ordinary} part of the cocycle will be determined by
\be
\nabla\nabla S_2{\big\vert} \approx -4 \nabla^\beta\bar \nabla_{\dal}G_a\, \nabla_\beta \bar \nabla^{\dal} G^a{\big\vert} + 2 \nabla^\alpha \nabla_\alpha R \,\bar \nabla_{\dal}\bar\nabla^{\dal} R^+{\big\vert}\label{ordS2}
\ee
The second term is well known, see \cite{WB}. We have $\nabla\nabla R{\big\vert}\approx -\frac 13 \ER$, so
\be
\nabla^\alpha \nabla_\alpha R\, \bar \nabla_{\dal}\bar\nabla^{\dal} R^+{\big\vert}\approx \frac 19 \ER^2\label{S21}
\ee
It remains for us to compute $\nabla_\beta\bar \nabla_{\dal}G_a{\big\vert}$. From (\ref{Rscomposto}) we can derive
$\bar \nabla_{\dal}\nabla_\beta G_a{\big\vert}$. On the other hand we have
\be
(\nabla_\beta\bar \nabla_{\dal}+\bar \nabla_{\dal}\nabla_\beta) G_a=R_{\dal \beta a}{}^b G_b-T_{\dal \beta}{}^B \nabla_BG_a=
-2 G_a \sigma^b{}_{\beta\dal} G_b -2i \sigma^b{}_{\beta\dal} \nabla_b G_a\approx 0\0
\ee
Therefore
\be
\nabla_\beta\bar \nabla_{\dal}G_a\approx -\bar \nabla_{\dal}\nabla_\beta G_a\label{DDGa}
\ee
Next, using the notation $\bar \nabla_{\dal}\nabla_{\al} G_{\beta\dbeta}= \sigma^a{}_{\beta\dbeta} \bar \nabla_{\dal}\nabla_{\al}G_a$, we introduce the following decomposition
\be\label{decomp}
\bar \nabla_{\dal}\nabla_{\al} G_{\beta\dbeta}= A_{(\al\beta)(\dal\dbeta)}+\epsilon_{\al\beta} \,B_{\dal\dbeta} +
\epsilon_{\dal\dbeta}\, C_{(\al\beta)}+ \epsilon_{\al\beta}\,\epsilon_{\dal\dbeta} \,D
\ee
Now we remark that (\ref{Rscomposto}) contains the part of $\bar \nabla_{\dal}\nabla_{\al} G_{\beta\dbeta}$ which is symmetric both in the couple $\al,\beta$ and $\dal,\dbeta$. After some lengthy but straightforward calculation
one can extract it and get
\be
A _{(\al\beta)(\dal\dbeta)}=-\frac 12 R_{abcd}\, (\sigma^{ab} \epsilon)_{\al\beta}\, (\epsilon \bar\sigma^{cd})_{\dgamma\ddelta}\label{A}
\ee
Next, contracting the decomposition (\ref{decomp}) with $\epsilon^{\beta\al}$ and using the first equation
in (\ref{constraints}) we get
\be
\epsilon^{\beta\al}\bar \nabla_{\dal}\nabla_{\al} G_{\beta\dbeta}=2\,B_{\dal\dbeta}+2 \epsilon_{\dal\dbeta}\,D= \nabla_{\dal}\nabla_{\dbeta}R^+\approx -\frac 12 \epsilon_{\dal\dbeta}\,\bar\nabla\bar \nabla R^+\label{eDDG}
\ee
A similar result one gets by contracting (\ref{decomp}) with $\epsilon^{\dal\beta}$.
We conclude that
\be
&&B_{(\dal\dbeta)}{\big\vert}\approx 0, \quad\quad C_{(\al\beta)}{\big\vert}\approx 0\0\\
&& D{\big\vert}\approx -\frac 14 \nabla\nabla R{\big\vert} \approx -\frac 14 \bar\nabla\bar\nabla R^+{\big\vert}\approx \frac 1{12} \ER\label{BCD}
\ee
The remaining computation is straightforward. We get
\be
\nabla\nabla \ES_2{\big\vert} &\approx& -4 \nabla^\beta\bar \nabla_{\dal}G_a\, \nabla_\beta \bar \nabla^{\dal} G^a{\big\vert} + 2 \nabla^\alpha \nabla_\alpha R \,\bar \nabla_{\dal}\bar\nabla^{\dal} R^+{\big\vert}\label{S2}\\
&\approx& \frac 49 \ER^2 -2\ER_{nm}\ER^{nm}+\frac 29 \ER^2= \frac 23 \ER^2-2\ER_{nm}\ER^{nm}\0
\ee
that is
\be
\Delta_\sigma^{(2)} = 4\int d^4x \, e\, \omega \bigl(\frac 3 \ER^2-2\ER_{nm}\ER^{nm}\bigr)\label{Delta2}
\ee
This is not the Gauss-Bonnet density, as one could have expected. But it is easy to recover it by means of a
linear combination of $\Delta_\sigma^{(1)}$ and $ \Delta_\sigma^{(2)}$:
\be 
\Delta_\sigma^{(1)} +\frac 12 \Delta_\sigma^{(2)} \approx
\int d^4x \, e \Bigl\{\omega\bigl( \ER_{nmkl} \ER^{nmkl}-4 \ER_{nm}\ER^{nm} + \ER^2\bigr) - \frac 12\, \alpha \, \epsilon^{nmlk}\ER_{nmpq}\ER_{lk}{}^{pq}\Bigr\}\label{GB}
\ee
which contains precisely the Gauss-Bonnet density\footnote{For an early appearance of the Gauss-Bonnet and Weyl density anomalies in supergravity see \cite{TN,CD2}.}.

{\it In conclusion $\Delta_\sigma^{(1)}$ corresponds to a multiplet of anomalies, whose first component is the Weyl density multiplied by $\omega$, accompanied by the Pontryagin density (the Delbourgo-Salam anomaly) multiplied by
$\alpha$. On the other hand $\Delta_\sigma^{(2)}$ does not contain the Pontryagin density and the part linear
in $\omega$ is a combination of the Weyl and Gauss-Bonnet density.}

\section{Non minimal supergravity}

In supergravity there is a freedom in imposing the torsion constraints. A convenient choice is in terms of the so-called `natural constraints'
\be 
&&T_{ab}{}^c=0,  \quad\quad
T_{\alpha\beta}{}^a=T_{\dot\alpha\dot\beta}{}^a=0,\quad\quad T_{\alpha\dot\beta}{}^a= 2i \sigma^a_{\alpha\dot\beta},\0\\
&&T_{\gamma}{}^{\dot \beta}{}_{\dot\alpha} =(n-1) \delta_{\dot\alpha}^{\dot\beta}\, T_\gamma, \quad\quad T^{\dot\gamma}{}{_\beta}{}^{\alpha} =(n-1) \delta^{\alpha}_{\beta} \,\bar T^{\dot\gamma}\label{naturalc}\\
&& T_{\gamma\beta}{}^\alpha= (n+1) (\delta^\alpha_\gamma\, T_\beta +\delta^\alpha_\beta \, T_\gamma ),\quad\quad  T^{\dot\gamma\dot\beta}{}_{\dot\alpha}= (n+1) (\delta_{\dot\alpha}^{\dot\gamma} \,\bar T^{\dot\beta} +\delta_{\dot\alpha}^{\dot\beta} \, \bar T^{\dot\gamma} )\0\\
&&T_{\gamma b}{}^a = 2n\, \delta_b^a \,T_\gamma, \quad\quad T^{\dot\gamma}{}_ b{}^a = 2n\, \delta_b^a\, \bar T^{\dot\gamma}\0
\ee
where $n$ is a numerical parameter and $T_\alpha,\bar T_{\dot\alpha}$ are new (conjugate) superfields in addition to those of minimal supergravity. The latter is obtained by setting $T_\alpha=0$. $T_\alpha,\bar T_{\dot\alpha}$ are U(1) connections. The $U(1)\times U(1)$ gauge symmetry was added with the purpose of enlarging the minimal supergravity model.
The solution for the Bianchi identities can be found in \cite{GGMW1}.
There are many significant changes with respect to the minimal model. For instance
$W_{\alpha\beta\gamma}$ and $R$ are not chiral anymore, but
\be 
&&(\bar \ED_{\dot\alpha} +(3n+1) \bar T_{\dot\alpha})W_{\alpha\beta\gamma}=0\label{nonchiral1}\\
&&(\bar \ED_{\dot\alpha} +2(n+1) \bar T_{\dot\alpha})R=0\label{nonchiral2}
\ee
where $\ED$ replaces $\nabla$ as covariant derivative\footnote{In principle there is no reason to use two different symbols for the covariant derivative, they denote the same covariant derivative in different settings. The use of two different symbols, however, will be instrumental in section 7.}.

A distinguished superfield is $S$ (and its conjugate $\bar S$), defined by
\be 
S= \ED^\alpha T_\alpha-(n+1) T^\al T_\al,\label{S}
\ee
which satisfies
\be
\ED_\al S=8T_\al R^+\label{del S}
\ee
The combination
\be 
Y=8R+2(n+1)\bar S\label{Y}
\ee 
is chiral, $\bar\ED^{\dal} Y=0$. The operator
\be
\Delta= \ED^\al \ED_\al -3(n+1) T^\al \ED_\al -Y\label{Delta}
\ee
projects a superfield without Lorentz indices to an antichiral superfield and
$\frac {\Delta}{Y}$ is a chiral projector.

The {\it non minimal model} for supergravity is obtained by further imposing the constraint
\be 
R=R^+=0\label{R=0}
\ee
with nonvanishing $T_\al$ and $\bar T_{\dal}$.

The non minimal supergravity has 20+20 degrees of freedom. The bosonic dofs are
those of the minimal model, excluding $R$ and $R^+$, plus 10 additional ones which can be identified with the lowest components of the superfields $S,\,\bar S$,$\, \bar\ED_{\dal}T_\al= c_{\al\dal}+id_{\al\dal}$ and $\bar\ED_{\al}T_{\dal}=- c_{\al\dal}+id_{\al\dal}$.
The additional fermionic dofs can be identified with the lowest components of $T_\al, \bar T_{\dal}$ and $\ED_\al \bar S,\, \bar\ED_{\dal} S$.

\subsection{Superconformal transformations in the non minimal model}

In the non minimal model there are transformations compatible with the constraints that correspond to local vierbein rescalings. We will refer to them generically as superconformal transformations. They are good candidates for superWeyl transformations (i.e, for supersymmetric extensions of the ordinary Weyl transformations) but, as we shall see, do not automatically correspond to them. They are expressed in terms of an arbitrary (complex) superfield $\Sigma$
\be 
&& \delta E_\al {}^M = -(2\bS-\Si)\, E_\al {}^M\0\\
&& \delta E_{\dal} {}^M = -(2\Si-\bS)\, E_{\dal} {}^M\0\\
&&\delta E_a{}^M=-(\Si+\bS)\, E_a{}^M+ \frac i2 \bar\sigma_a^{\dbeta\beta}\, \bar \ED_{\dbeta}\left( \bS-\frac {3n-1}{3n+1} \Si\right)\, E_\beta{}^M\0\\
&&\quad\quad\quad\quad +\frac i2 \bar\sigma_a^{\beta\dbeta} \ED_{\beta}\left( \Si-\frac {3n-1}{3n+1} \bS\right) E_{\dbeta}{}^M\0\\
&&\delta T_{\al}= -(2\bS-\Si)\, T_{\al}+\frac 3{3n+1} \ED_{\al}\bS\label{nonmin}\\
&&\delta W_{\al\beta\gamma} =-3\Si\, W_{\al\beta\gamma}\0\\
&& \delta G_a= -(\Si+\bS)\,G_a +i\ED_a (\bS-\Si) +\frac 13 \bar\sigma_a^{\dal\al}
\left( T_{\al}\bar \ED_{\dal}\bS-\bar T_{\dal}\ED_{\al} \Si\right)\0\\
&&\quad\quad\quad -\frac {3n-1}{3(3n+1)} \bar\sigma_a^{\dal\al} \left( T_{\al}\bar \ED_{\dal}\Si-\bar T_{\dal}\ED_{\al} \bS\right)\0\\
&&\delta R^+ = -2(2\bS -\Si)R^+\0\\
&&\quad\quad\quad +\frac 1{4(3n+1)}\left(\ED^{\al}\ED_{\al} +(n+1) T^{\al}\ED_{\al} \right) \left[3n(\bS-\Si)-(\bS+\Si)\right]\0
\ee
From (\ref{R=0}) and (\ref{nonmin}) we see that the superfield $\Si$ is constrained by the linear condition
\be
\left(\ED^{\al}\ED_{\al} +(n+1) T^{\al}\ED_{\al} \right) \left[3n(\bS-\Si)-(\bS+\Si)\right]=0\label{nonminconst}
\ee

\subsection{Cocycles in non minimal SUGRA}

It the non minimal model it is easy to construct an invariant (0-cocycle)
\be
I^{(1)}_{n.m.}= \int_{x,\theta} E W^{\alpha\beta\gamma}W_{\alpha\beta\gamma} \frac{\bar T_{\dot\alpha}\bar T^{\dot\alpha}}{\bar S^2}+ h.c.\label{inv1}
\ee
and a 1-cocycle
\be
\Delta^{(1)}_{n.m.}= \int_{x,\theta} E\, \Sigma\, W^{\alpha\beta\gamma}W_{\alpha\beta\gamma} \frac{\bar T_{\dot\alpha}\bar T^{\dot\alpha}}{\bar S^2}+ h.c.\label{Delta1nm}
\ee
It is easy to prove that $\delta I_{n.m.}^{(1)}=0=\delta \Delta^{(1)}_{n.m.}$ for any $n$. To this end the condition (\ref{nonminconst}) is inessential. If $R\neq 0$ this is not true anymore. 

The construction of a second cocycle corresponding to $\Delta_\sigma^{(2)}$  above, is not as straightforward and will be postponed to section 7, after the technique of mapping between different supergravity models has been introduced.

\section{The 16+16 new minimal model}

One way to define the new minimal model is to introduce a 2-superform $B_{AB}$ and impose natural constraints on its supercurvature. In this way we obtain a 16+16 model. 
The independent bosonic dofs are the vierbein, the lowest component of $S, \bar S$, $c_{\al\dal}$
and $G_{\al\dal}$ (the components of $d_{\al\dal}$ are not independent in this model). The fermionic degrees of freedom are, beside the gravitino field, the lowest components of $T_\al,\bar T_{\dal}$ and $\ED_\al \bar S, \ED_{\dal}S$. The new dofs (with respect to the minimal model) are linked to the mode contained in $B_{ab}$. In new minimal supergravity the range of the parameter $n$  is limited to $n>0$ and $n<-\frac 13$.

In practice this means that
\be
T_\al = \ED_{\al} \psi, \quad\quad \quad T_{\dal}= \ED_{\dal}\psi\label{newminconst}
\ee
where $\psi$ is a (dimensionless) real superfield. The transformations corresponding to (\ref{nonmin}) on $\psi$ are
\be 
\delta \psi = \frac 3{3n+1} (\bS-\bar \Lambda)= \frac 3{3n+1} (\Si - \Lambda)\equiv\frac 3{3n+1}L\label{psitransf}
\ee
where $\Lambda (\bar\Lambda)$ is an arbitrary chiral (antichiral) superfield, and $L$ is a real (vector) superfield. As a consequence the transformations (\ref{nonmin}), compatible with the constraints, for the surviving superfields take the form:
\be 
&& \delta E_\al {}^M = -(L+2\bL-\La)\, E_\al {}^M\0\\
&& \delta E_{\dal} {}^M = -(L+2\La-\bL)\, E_{\dal} {}^M\0\\
&&\delta T_{\al}= -(L+2\bL-\La)\, T_{\al}+\frac 3{3n+1} \ED_{\al}L\label{newmin}\\
&&\delta W_{\al\beta\gamma} =-3(L+\La)\, W_{\al\beta\gamma}\0\\
&& \delta G_{\al\dal}= -(2L+\La+\bL)\,G_{\al\dal}+i\ED_ {\al\dal} (\bL-\La) -\frac 23
\left( T_{\al}\bar \ED_{\dal}\bL-\bar T_{\dal}\ED_{\al} \La\right)\0\\
&&+\frac {2(3n-1)}{3(3n+1)}\left(T_{\al}\bar \ED_{\dal} L-\bar T_{\dal}\ED_{\al} L\right)\0\\
&& \delta S= -2(L+2\bL-\La) S +4 \ED^\al (L+\La) T_\al- \frac {21n+5}{3n+1}\ED^\al  L\, T_\al+ \frac 3{3n+1}  \ED^\al \ED_\al L \0 
\ee
and (\ref{nonminconst}) becomes
\be
\left(\ED^{\al}\ED_{\al} +(n+1)T^{\al}\ED_{\al} \right)(L+(3n+1)\La)=0 \label{linconst}
\ee

\subsection{Cocycles in new minimal 16+16 SUGRA}

As in nonminimal SUGRA it is easy to construct an invariant 
\be
I^{(1)}_{new}= \int_{x,\theta} E W^{\alpha\beta\gamma}W_{\alpha\beta\gamma} \frac{\bar T_{\dot\alpha}\bar T^{\dot\alpha}}{\bar S^2}+h.c.\label{inv1new}
\ee
and a 1-cocycle
\be
\Delta^{(1)}_{new}= \int_{x,\theta} E\, (L+\Lambda) W^{\alpha\beta\gamma}W_{\alpha\beta\gamma} \frac{\bar T_{\dot\alpha}\bar T^{\dot\alpha}}{\bar S^2}+h.c.\label{Delta1new}
\ee
It is easy to prove that $\delta I_{new}^{(1)}=0=\delta \Delta^{(1)}_{new}$. Once again we don't need (\ref{linconst}) to prove this. On the other hand it is not easy to construct a cocycle similar to $\Delta^{(2)}$, i.e. quadratic in the superfield $G_a$, which, after translation to component form, leads to the Gauss-Bonnet density.

\section{Reduction to component form}
\label{app:redux}

In the following analysis the reduction of the cocycles to ordinary form will play a major role.
Thus the purpose of this section is to outline the procedure to derive the component form of the cocycles in nonminimal and new minimal supergravities, as we have done in section 2 for the minimal supergravity anomalies. The operator, \cite{GGMW1},
\be
\bar \Delta =\bar \ED_{\dal}\bar\ED^{\dal}-3(n+1) \bar T_{\dal} \bar\ED^{\dal}-  2(n+1)\bar S \label{barDelta}
\ee
projects a generic superfield to a chiral superfield. Let $U$ be a superfield without Lorentz indices. It is not hard to see that (see \cite{GGMW1})
\be
\int_{x,\theta} E\,  U=\frac 1{4n}\int_{x,\theta} E \,\bar \Delta \left(e^{-\bar\Omega}\, \frac U{\bar S}\right)\label{real1}
\ee
where $\bar \Omega = 2(3n+1) \frac {\bar T \bar T}{\bar S}$. 
Therefore, introducing the appropriate chiral density ${\cal E}$, \cite{GMW}, we can write
\be
\int_{x,\theta} E\,  U= \frac {1}{4n} \int d^4x\int d^2\Theta\, {\cal E}\,\bar \Delta  \left(U \, e^{\bar \Omega}\right)\label{real2}
\ee
For instance, when $\Sigma$ is a chiral superfield the anomaly (\ref{Delta1nm})  can be written
\be
\Delta^{(1)}_{\Sigma}= \frac 1{4n} \int d^4x\int d^2\Theta \,{\cal E}\,  \bar \Delta \Big{[}\Sigma\, W^{\alpha\beta\gamma}W_{\alpha\beta\gamma} \frac{\bar T_{\dot\alpha}\bar T^{\dot\alpha}}{\bar S^2}\Big{]}+h.c.\label{real3}
\ee
When $\Sigma$ is not chiral there is in the RHS an additional term which, however, is irrelevant for the following considerations and so will be dropped. In a similar way we can deal with (\ref{Delta1new}). After some algebra we have in particular
\be
\Delta^{(1)}_{\Sigma}=-\frac 1{4n} \int d^4x \int d^2\Theta\, {\cal E}\,\left( \Sigma  \, W^{\al\beta\gamma}W_{\al\beta\gamma}+2\, W^{\al\beta\gamma}W_{\al\beta\gamma} \frac {\bar T_{\dal} \ED^{\dal} \Sigma}{ \bar S}\right)+h.c.\label{realfinal}
\ee
Therefore, proceeding as in section 5, $\Delta^{(1)}_{\Sigma}$ in components becomes
(we disregard a multiplicative factor)
\be
\Delta^{(1)}_{\Sigma}\approx  \int d^4x \, e\, \Bigl[\Sigma\,\ED^\alpha W^{\beta\gamma\delta}
\,\ED_\alpha W_{\beta\gamma\delta}{\big\vert}+h.c.\Bigr]\label{real4}
\ee
Now (\ref{dWsymm},\ref{dbarWsymm}) remain valid in the non minimal SUGRA, but (\ref{symmetrized}) is replaced by a far more complicated equation, so that (\ref{approx}) has to be re-demonstrated. This is not trivial, but can be done (see Appendix A). Thus we can conclude that, up to a multiplicative factor,
\be
\Delta^{(1)}_{\Sigma}\approx\frac 14 \int d^4x \, e \Bigl\{\omega\biggl( \ER_{nmkl} \ER^{nmkl}-2 \ER_{nm}\ER^{nm} +\frac 13 \ER^2\biggr) - \frac 12\, \alpha \, \epsilon^{nmlk}\ER_{nmpq}\ER_{lk}{}^{pq}\Bigr\}\label{Weyl+DS}
\ee
where $\omega+i\alpha$ is the lowest component of the superfield $\Sigma$. In this case too 
$\omega$ corresponds to the ordinary Weyl rescaling, while $\alpha$ is the parameter of a chiral transformation.

The same reduction to ordinary form holds also for (\ref{Delta1new}). In this case $\omega+i\alpha$ is the first component of $L+\La$.

At this point it is worth making a comment on the (apparent) singularity of expressions such as (\ref{inv1},\ref{Delta1nm},\ref{inv1new},\ref{Delta1new}). For instance, the cocycle (\ref{Delta1nm}), written in terms of superfields has a nonlocal or singular aspect; but one must reflect on the fact that it is nothing but the supersymmetrization of (\ref{Weyl+DS}), which is local. Therefore also (\ref{Delta1nm}), when expressed in terms of components fields must be local (although it may be non-polynomial if dimensionless prepotentials have to be introduced). 
The question remains open of whether non-singularity can be be made manifest already at the superfield level. In \cite{GMW} it was noted that in some cases this is indeed possible by means of opportune field redefinitions.

The scheme outlined in this section is general and will be applied to all the cocycles we will come across.

\section{Mapping formulas between different supergravity models}

A cocycle similar to $\Delta^{(2)}_{\sigma}$ (i.e. quadratic in $G_a$) is hard to construct with ordinary means (i.e. with a polynomial cohomological analysis) for ${\mathfrak W}$. For this we have to resort to a mapping between different supergravity models. This mapping was outlined in \cite{GMW,GGMW2} and brought to a more explicit form in \cite{Kuzenko}. The latter reference is based on different torsion constraints with respect to (\ref{naturalc}). Therefore we have to rederive new appropriate mapping formulas.
 
Various different models of supergravity are defined by making a definite choice of the torsion constraints and, after such a choice, by identifying the dynamical degrees of freedom. This is the way minimal, nonminimal and new minimal models were introduced. However it is possible to transform the choices of constraints into one another by means of suitable linear transformation of the supervierbein and the superconnection, \cite{GGMW1,Kuzenko}:
\be
E'{}_M{}^A= E_M{}^B X_B{}^A, \quad\quad E'{}_A{}^M= X^{-1} {}_A{}^B E_B{}^M,\quad\quad
\Phi'{}_{MA}{}^B=\Phi_{MA}{}^B +\chi_{MA}{}^B\label{EXE}
\ee
For instance, if we want to pass from a set of unprimed constraints to primed ones the required transformations are as follows 
\be 
&&E'_\al =U\,E_\al, \quad\quad E^{' \dal}= \bar U \,E^{\dal},\quad\quad E'= U^{-2}\bar U^{-2} E\label{minnonmin1}\\
&& E'_{\al\dal} = U\bar U \, E_{\al\dal} +i\frac {U\bar U}{3n+1}\left(E_{\dal}{}^M \del_M \ln\left(\frac {U^{n+1}}{\bar U^{n-1}}\right) E_\al  + E_\al{}^M \del_M \ln \left( \frac {\bar U^{n+1}}{U^{n-1}}\right) E_{\dal}\right)\label{minnonmin2}
\ee
where $U$ is a suitable expression of the superfields. Moreover
\be 
T'_\al &=& U T_\al -\frac 1{6n+2} \, \ED'_\al \ln \left(\bar U^{2}U^4 \right) \label{Tal1}\\
\Phi'_{\al\beta\gamma} &=& U\, \Phi_{\al\beta\gamma} -\frac 1{3n+1} \left(\epsilon_{\al\gamma}
\ED'_\beta  +\epsilon_{\al\beta}
\ED'_\gamma \right) \ln \left(\frac {U^{n-1}}{\bar U^{n+1}}\right)\label{Tal2}\\
W'_{\al\beta\gamma} &=& U\,\bar U^2 W_{\al\beta\gamma}\label{Tal3}\\
8R' + 2(n+1) \bar S'&=& -\left(\bar \ED'_{\dal}\bar\ED^{'\dal} -3(n+1) \bar T_{\dal}\bar\ED^{'\dal} -8R-2(n+1) \bar S\right) \bar U ^2\label{Tal4}
\ee
where $\ED'$ denotes the covariant derivative in the primed system, together with the conjugate relations. The analogous transformation for the $G_a$ superfield is more complicated:
\be 
G'_{\al\dal}&=& U\bar U\left( G_{\al\dal} -\frac i3 \ED'_{\al\dal} \ln \frac U{\bar U}+\frac 1{(3n+1)^2}\bar \ED'_{\dal} \ln \frac {U^{n+1}}{\bar U^{n-1}} \ED'_{\al} \ln \frac {\bar U^{n+1}}{U^{n-1}}\right.\0\\
&&+\frac 1{3(3n+1)} \bar\ED'_{\dal} \ln \frac {U^{n+1}}{\bar U^{n-1}}\ED'_\al \frac U{\bar U}+ \frac 1{3(3n+1)} \ED'_{\al} \ln \frac {\bar U^{n+1}}{U^{n-1}}\bar\ED'_{\dal} \frac U{\bar U}\0\\
&&\left. + \frac 2{3(3n+1)} \ED'_{\al} \ln \frac {\bar U^{n+1}}{U^{n-1}} \bar T_{\dal} -  \frac 2{3(3n+1)} \bar\ED'_{\dal} \ln \frac {U^{n+1}}{\bar U^{n-1}} \, T_\al\right)\label{Gaaldal}
\ee 
These formulas can be inverted. To this end we have to replace $U$ with $U^{-1}$ everywhere, replace the primed quantities with unprimed ones in the LHS, and the unprimed with the primed ones in RHS; in this case the covariant derivatives on the RHS are the primed ones\footnote{For more details on these transformations, see \cite{Giaccari}.}.

For instance, if we want to pass from the minimal to the nonminimal constraints we have to choose
\be
U= \exp \left[ 2(3n+1) \left(\frac {\bar \psi}6-\frac {\psi} 3\right)\right]\label{Udef}
\ee
$\psi$ is a `prepotential' such that $T_\al = \ED_\al \psi$ and $\bar T_{\dal} =\bar \ED_{\dal} \bar \psi$. Of course if we wish to pass from the nonminimal to the minimal constraints we have simply to use the same formulas with inverted $U$. 

One can verify that
\be 
(\bar \nabla_{\dal} \bar \nabla^{\dal} -8R) \bar U^2= -2(n+1) \bar S\label{doblenabla}
\ee
We recall that $\nabla$ denotes specifically the covariant derivative in minimal supergravity.

Let us consider next the superconformal transformations. We wish to compare the transformations (\ref{superconf},\ref{superfieldconf}) with (\ref{nonmin}). Given the transformation of $T_\al$ and $T_\al = \ED_\al \psi$,  we can assume that $\psi, \bar \psi$ transform as follows
\be
\delta \psi =\frac 3{3n+1} (\bar \Sigma -\bar\sigma),\quad\quad
\delta \bar \psi =\frac 3{3n+1} ( \Sigma -\sigma), \label{deltapsi}
\ee
where $\sigma$ is an arbitrary chiral superfield. Taking the variation of both sides of 
(\ref{minnonmin1}) and applying (\ref{deltapsi}) we can easily see that we can identify the $\sigma$ superfield in (\ref{deltapsi}) with the $\sigma$ in (\ref{superconf}). The same is easily done also for (\ref{Tal3}). The transformation of (\ref{Tal4}) is more complicated. We first derive, using (\ref{Tal2}), 
\be 
\nabla^\al\nabla_\al \Phi = U^{-2} \left( \ED\ED \Phi -\frac 43 (3n+1) \ED^\al \bar T \ED_\al \Phi + \frac {15n-1}3 T^\al \ED_\al \Phi\right)\label{nablanablaphi}
\ee
for any scalar superfield $\Phi$. Inverting (\ref{Tal4}) we can write
\be 
-8 R^+ = \left(\ED^\al \ED_\al- 3(n+1) T^\al \ED_\al -2(n+1)S\right) U^{-2}\label{Tal4'}
\ee
The LHS represents $R$ in the minimal model, while the RHS refers to the nonminimal one.
Taking the variation of both sides and using (\ref{deltapsi},\ref{nablanablaphi}), one can show that
\be
\delta R^+= - 2(2\bar \sigma-\sigma) R^+-\frac 14 \nabla\nabla \sigma\label{deltaR+}
\ee
This is identical to the transformation of $R^+$ in the minimal model, (\ref{superfieldconf}).

We can do the same with $G_a$. Taking the variation of LHS and RHS of the inverted eq.(\ref{Gaaldal}), and using 
\be 
i\nabla_{\al\dal}(\bar \sigma-\sigma)  =U^{-1}\bar U^{-1} \left( i\ED_{\al\dal}(\bar \sigma-\sigma) -\frac 1{3n+1} 
\bar \ED_{\dal}\ln \frac {U^{n+1}}{\bar U^{n-1}}\,\ED_\al \sigma + \frac 1{3n+1} \ED_{\al}\ln \frac {\bar U^{n+1}}{U^{n-1}}\, \bar \ED_{\dal}\bar \sigma\right)\0
\ee
one finds
\be
\delta G_{\al\dal}= -(\sigma+\bar \sigma) G_{\al\dal} +i \nabla_{\al\dal}(\bar \sigma-\sigma)\label{DeltaGaldal}
\ee
as expected.

Therefore (\ref{deltapsi}) connects the superconformal transformations of the minimal and nonminimal models. It is however useful to consider this passage in two steps.
Let us split $U$ in (\ref{Udef}) as follows:
\be 
U=U_c U_n ,\quad\quad U_c=e^{X-2\bar X},\quad\quad U_n = e^{\frac {\Omega}3-\frac {\bar \Omega}6}\label{Usplit}
\ee
where 
\be 
X=\frac 13 (3n+1) \bar \psi+\frac {\bar \Omega}6,\quad\quad
\bar X= \frac 13 (3n+1) \psi+\frac { \Omega}6\label{XbarX}
\ee
Recall that $\Omega =2(3n+1)\frac { T^{\al} T_\al}{S}$ and $T_\al =\ED_\al \psi$, etc.
It follows that $X$ is a chiral and $\bar X$ an antichiral superfield. Moreover  
$U_c \bar U_c^2= e^{-3 X}$ is chiral and $\bar U_c U_c^2= e^{-3\bar X}$  is antichiral.
Operating on the superfields according to (\ref{minnonmin1},\ref{Tal1},\ref{Tal3}) we see that, for instance $T_\al=0$ is mapped to $T_\al=0$ by the transformation induced by $U_c$, i.e. after such transformation the model is still minimal supergravity.

For later use we remark that (see also \cite{GMW})
\be 
\delta \bar\Omega = \Gamma_\Sigma-6\Sigma, \quad\quad \quad\Gamma_\Sigma = -\frac 3{3n+1} \bar \Delta\left(\frac {\bar \Omega \Sigma}{\bar S}\right)\label{Gamma}
\ee
where $\bar \Delta$ is the chiral projector. By repeating the previous verifications one can see that $\Gamma_\Sigma$ is an intermediate step between $\sigma$ and $\Sigma$. The important property of $\Gamma_\Sigma$ is that it is chiral, but expressed in terms of the nonminimal superfields. Moreover it is consistent with the nonminimal transformation properties and, in particular,
$\delta \Gamma_\Sigma=0$. In parallel with (\ref{Gamma}) we have of course the conjugate formulas.

Analogous things hold if we replace the non minimal with the new minimal model. In this case of course we have to set $\psi=\bar\psi$ and the appropriate transformations are (\ref{psitransf},\ref{newmin}). It is easy to see that the above superfield redefinitions connect the minimal supergravity transformations with (\ref{newmin}). Also in this case we have an intermediate step which will turn out instrumental later on. In this case we have
\be 
\delta \bar\Omega = \Gamma_{L+\La}-6(L+\Lambda), \quad\quad \quad\Gamma_{L+\La} = -\frac 3{3n+1} \bar \Delta\left(\frac {\bar \Omega (L+\La) }{\bar S}\right)=\Gamma_L+6\Lambda\label{GammaL}
\ee
where $\Gamma_L$ is chiral.

All this means one important thing: the possibility to construct invariants and cocycles of any supergravity model starting from the invariants and cocycles of a fixed one, for instance the minimal supergravity (such an idea is present in \cite{Brandt1}).

\section{Cocycles from minimal supergravity}

We are now ready to construct the cocycles form those of minimal supergravity. The idea is very simple. We start from the cocycles of minimal supergravity and replace the superfields of the latter with the formulas of the previous subsection expressing them in terms of the superfields of other models. Since all the symmetry operations are coherent, the resulting expressions must also be cocycles. The invariants are a subcase of the discussion for 1-cocycles, thus in the sequel we explicitly deal only with the latter. We will consider first the new minimal case.

\subsection{From minimal to nonminimal cocycles}

Let us start from
$\Delta_\sigma ^{(1)}$. All the superfields therein must be expressed in terms of the new superfields. It is convenient to proceed in two steps, as just outlined. In the first step it is mapped to
\be 
\int_{x,\theta} \frac E{-8R}\sigma\, WW+ h.c.= \int_{x,\theta} E'\Gamma_\Sigma \, \frac {W'W'}{\bar U_c^2(\nabla'_{\dal}\nabla^{'\dal}-8R'){\bar U}_c^{-2}}+h.c.\label{Delta1sigma1}
\ee
where $WW$ is a compact notation for $W^{\al\beta\gamma}W_{\al\beta\gamma}$ and primes denote the superfields in the new representation (which still corresponds to minimal supergravity). We recall that $\nabla'_{\dal}\nabla^{'\dal}-8R'$ projects to a chiral superfield. Therefore we can write
\be 
\int_{x,\theta} \frac {E\, \sigma}{-8R}\, WW&=& \int_{x,\theta} \frac {E'}{-8R'} \Gamma_\Sigma
(\nabla'_{\dal}\nabla^{'\dal}-8R') \left( \frac{W'W'\bar U_c^{-2}}{(\nabla'_{\dal}\nabla^{'\dal}-8R'){\bar U}_c^{-2}}\right)\\
&=&
\int_{x,\theta} \frac {E'}{-8R'}\Gamma_\Sigma \, W'W'\label{Delta1sigma2}
\ee
Now we complete the passage to the nonminimal model by performing the $U_n$ transformation.
This means
\be
\int_{x,\theta} \frac {E'}{-8R'}\Gamma_\Sigma \, W'W'= \int_{x,\theta} {E''}\Gamma_\Sigma\,
\frac {W''W''}{\bar U_n^2\bar \Delta''\bar U_n^{-2}}\label{Delta1sigma3}
\ee
where $\Delta'' =\ED''\ED''-3(n+1){ T''}^\al\ED''_\al-2(n+1)S''$ is the antichiral projector in the nonminimal model (endpoint of the overall transformation). For simplicity, from now on, we drop primes, understanding that we are operating in the nonminimal model.  

Next we use the identity, demonstrated in \cite{GGMW1} by partial integration,
\be 
4 n \int_{x,\theta} E \, e^{\bar \Omega} U = \int_{x,\theta} E\, \frac {\Phi}{\bar S}\label{GGMW}
\ee
where $U$ is any superfield expression without Lorentz indices and $\Phi=\bar \Delta U$.
Applying this identity with $U=\Sigma \frac {e^{-\bar\Omega} WW\, \bar U^{-2}}{\bar \Delta\bar U^{-2}}$ we get
\be
  \int_{x,\theta} E\,\Gamma_\Sigma\,  \frac { WW}{ \bar U^2\,\bar \Delta\bar U^{-2}}
= \frac 1{4n} \int_{x,\theta}\frac  E {\bar S}\,\Gamma_\Sigma\, e^{-\bar\Omega} WW\label{Delta1sigma4}
\ee
Applying (\ref{GGMW}) again with $U= \Sigma e^{-\bar \Omega} \frac {WW}{\bar S}$, so that
$\Phi=-2(n+1) \Gamma_\Sigma e^{-\bar \Omega} {WW}$, we obtain
\be 
\frac 1{4n}  \int_{x,\theta}\frac  E {\bar S}\Gamma_\Sigma\, e^{-\bar\Omega} WW= -\frac 1{2(n+1)}
 \int_{x,\theta}\frac  E {\bar S}\,\Gamma_\Sigma\, WW\label{Delta1sigma5}
\ee
Replacing now the explicit expression of $\Gamma_\Sigma$, (\ref{Gamma}), and integrating by parts, we find that
$\Delta^{(1)}_\sigma$ is mapped to
\be
3\frac {5n+1}{n+1} \int_{x,\theta}\frac  E {\bar S^2} \Sigma\, \bar T_{\dal}\bar T^\al\,
W^{\al\beta\gamma}W_{\al\beta\gamma} +h.c.
\ee
which is proportional to the already obtained cocycle $\Delta^{(1)}_{n.m.}$, (\ref{Delta1nm}).  
The second cocycle is readily constructed in the same way:
\be
\Delta_{\Sigma}^{(2)}&=& \int_{x,\theta} E' (\Gamma_\Sigma+\bar \Gamma_\Sigma)U^2 \bar U^2\left( G_a(G',T',U)  G^a(G',T',U)+2 R(\bar S',\bar T',U)\, R^+(S',T',U)\right) \0\\
&=&c\, \int_{x,\theta} E' (\Sigma+\bar \Sigma) (G_a' G^{'a}+\ldots)\ \label{Delta2sigma}
\ee
after repeated partial integrations. $G_a(G',T',U)$ is given by the inverted (\ref{Gaaldal}), while $R^+(S,T,U)$ is given by (\ref{Tal4'}). $c$ is a suitable number. By construction $ \Delta_{\Sigma}^{(2)}$ satisfies the consistency conditions with generic $\Sigma$. Its ordinary form is the same as $\Delta_{\sigma}^{(2)}$
in section 2.

\subsection{From minimal to new minimal cocycles}

Let us start again with $\Delta^{(1)}_\sigma$. Proceeding as above with the relevant new formulas outlined at the end of the previous section we get
\be 
\Delta^{(1)}_\sigma&=&\int_{x,\theta} \frac {E}{-8R} \sigma WW+h.c.\0\\
&=&  \int_{x,\theta} E'\Gamma_{L+\La} \frac{W'W'}{\bar U^2 \left(\bar\ED\bar\ED -3(n+1)\bar T'_{\dal} \bar \ED-2(n+1)\bar S'\right) \bar U^{-2}}+h.c.\equiv\tilde\Delta^{(1)}_{L+\Lambda} \label{Delta1prime}
\ee
where primed superfields refers to new minimal supergravity. From now on we drop primes, understanding that all the superfields are in the new minimal supergravity. Working out the derivatives in (\ref{Delta1prime}) we get 
\be 
\tilde\Delta^{(1)}_{L+\Lambda}= -\frac 34 \int_{x,\theta}\frac E{\bar S}\, \Gamma_{L+\La}\, WW\left( 1- \frac 23 (3n+1) \frac{\bar T_{\dal}\bar T^{\dal}}{\bar S}\right)+h.c.\label{tildeDelta1}
\ee
This is not (\ref{Delta1new}) yet, as we would have expected. However, using (\ref{real1})  
and integrating by parts the spinor derivatives contained in $\Gamma_{L+\La}$, as we have done above for the nonminimal case, one easily finds that $\tilde\Delta^{(1)}_{L+\Lambda}$
is proportional to (\ref{Delta1new}).

Let us come now to the second cocycle. As above we start from the minimal cocycle $\Delta_\sigma^{(2)}$ and transform the superfields according to (\ref{Tal4'}) and (\ref{Gaaldal}). We  get
\be 
\Delta_\sigma^{(2)}&=& \int_{x,\theta}  {E}(\sigma+\bar\sigma) (G_a G^a+ 2RR^+)\0\\
&=& \int_{x,\theta} {E'}(\Gamma_{L+\La}+h.c.) \left(-\frac 12 \left(G'_{\al\dal}+\frac 49 T_\al\bar T_{\dal}\right) \left(G^{'\al\dal}+\frac 49 T^\al\bar T^{\dal}\right)\right.\0\\
&&+\left. 2\left(\frac 16 S-\frac n3(3n+1)T^\al  T_\al\right) \left(\frac 16 \bar S-\frac n3(3n+1)\bar T_{\dal} \bar T^{\dal})\right) \right)\equiv\Delta^{(2)}_{L+\Lambda}\label{Delta2prime}
\ee
where superfields and covariant derivatives in the RHS are new minimal superfields. Of course since nothing has changed concerning the metric, the ordinary form of $\Delta^{(2)}_\Lambda$ is the same as the ordinary form of $\Delta^{(2)}_\sigma$, computed in section 2.

\section{Conclusions}

In this paper we have determined the possible trace anomalies in the new minimal supergravity as well as in the non minimal one. There are in all cases two independent nontrivial cocycles whose densities are given by the square Weyl tensor and by the Gauss-Bonnet density, respectively.  
 
Concerning the Pontryagin density, it appears in the anomaly supermultiplets only in the form of chiral anomaly (Delbourgo-Salam anomaly), but never in the form of trace anomaly.

At this point we must clarify the question of whether the cocycles we have found in nonminimal and new minimal supergravities are the only ones. In this paper we have not done a systematic search of such nontrivial cocycles in the nonminimal and new minimal case, the reason being that when a dimensionless field like $\psi$ and $\bar \psi$ are present in a theory a polynomial analysis is not sufficient (and a non-polynomial one is of course very complicated). But we can argue as follows: consider a nontrivial cocycle in nonminimal or new minimal supergravity; it can be mapped to a minimal cocycle which either vanishes or coincides with the ones classified in \cite{BPT}. There is no other possibility because in minimal supergravity there are no dimensionless superfields (apart from the vielbein) and the polynomial analysis carried out in \cite{BPT} is sufficient to identify all cocycles. We conclude that the nonminimal and new minimal nontrivial cocycles, which reduce in the ordinary form to a nonvanishing expression, correspond to $\Delta_\sigma^{(1)}$ and  $\Delta_\sigma^{(2)}$ in minimal supergravity.

Finally we would like to make a comment on an aspect of our results that could raise at first sight some perplexity. Although one cannot claim the previous results to be a theorem, they nevertheless point in the direction of the non-existence of a supersymmetric anomaly 
multiplet that has, as its e.m. tensor trace component, the Pontryagin density. On the other hand we know systems with chiral fermions that at first sight can be supersymmetrized and coupled to supergravity. In such system we expect the trace of the e.m. tensor at one loop to contain 
the Pontryagin density, \cite{BG2}; thus why couldn't we have an anomaly multiplet that contains as trace component the Pontryagin density? The point is that in such a chiral case there can exist an obstruction to that, as we try to explain next. Suppose that the e.m. tensor of a system like the one just mentioned, has, at one loop, an integrated nonvanishing trace  $\Delta_\omega^{(P)}$, containing a term given by $\omega$ multiplied by the Pontryagin density. We cannot expect, in general this term to be supersymmetric. On the contrary, denoting by $\epsilon$ the supersymmetric local parameter we expect there to exist a partner cocycle $ \Delta_\epsilon^{(P)}$ such that
\be
\delta_\omega \Delta_\omega^{(P)}=0, \quad\quad \delta_\epsilon \Delta_\omega^{(P)}+
\delta_\omega \Delta_\epsilon^{(P)}=0,\quad\quad \delta_\epsilon \Delta_\epsilon^{(P)}=0
\label{deltaodeltae}
\ee
The cocycle $\Delta_\epsilon^{(P)}$ to our best knowledge has not yet been computed in supergravity. So we have to rely on plausibility arguments.
There are two possibilities: it might happen that $ \Delta_\epsilon^{(P)}$ is trivial, i.e. $\Delta_\epsilon^{(P)}=\delta_\epsilon {\cal C}^{(P)}$, so that (\ref{deltaodeltae}) implies that $\delta_\epsilon(\Delta_\omega^{(P)}- \delta_\omega {\cal C}^{(P)})=0$. The end result would be a supersymmetric Weyl cocycle. This is, for instance, what happens for the chiral ABJ anomaly in rigid supersymmetry, where the supersymmetric partner of the usual chiral anomaly must be trivial, \cite{PS}, and, precisely as above, the chiral anomaly can be cast in supersymmetric form, see \cite{GGS}. 

The second possibility is that no such counterterm ${\cal C}^{(P)}$ exists, in which case the cocycle $ \Delta_\epsilon^{(P)}$ is nontrivial and there is no possibility to supersymmetrize
$\Delta_\omega^{(P)}$. This seems to be the case for the chiral ABJ anomaly in the presence of local supersymmetry, \cite{BPT2}. And this may be the case also for $\Delta_\omega^{(P)}$, which would explain the origin of our inability to find a Weyl cocycle containing the Pontryagin form in the first position (trace anomaly) in terms of superfield\footnote{A plausible explanation for the difference between local and global supersymmetry is that the nontrivial part of $ \Delta_\epsilon^{(P)}$ may be an integral of $\epsilon$ multiplied by (as it often happens) a total derivative; if $\epsilon$ is a (generic) local parameter $\Delta_\epsilon^{(P)}$ is a nonvanishing nontrivial cocycle, but it vanishes it $\epsilon$ is constant.}. In both cases the origin of the obstruction is the same, i.e. the nontrivial breaking of local supersymmetry. In turn this would explain why a supersymmetry preserving regularization has never been found in such types of systems. Such converging arguments seem to nicely fit together and close the circle.

\acknowledgments 

L.B would like to thank Taichiro Kugo, Jorge Russo and Adam Schwimmer for discussions and suggestions. We would like to thank Friedemann Brandt and Yu Nakayama for a useful exchange of messages. We would like to thank the referee of this paper for valuable criticisms and remarks. The work of L.B. and S.G. was supported in part by the MIUR-PRIN contract 2009-KHZKRX.

\vskip 1cm 
\section*{Appendix}
\appendix

\section{Reduction formulae}

In this appendix we collect from (\cite{GGMW1}) the formulas that are needed to reduce superfield expressions to component form. The equations below are not complete, they contain only the terms essential to recover the ordinary parts of the expressions (that is only the parts that survive once all the fields except the metric are disregarded). The complete form can be found in  (\cite{GGMW1}), or in \cite{WB} for the minimal model. The first formula when
evaluated at $\theta=\bar \theta=0$, connects the Riemann curvature to specific superfield  components and it is basic for reducing cocycles to ordinary form
\be
&&\sigma_{\al\dal}^a \sigma^b_{\beta\dbeta}\sigma^c_{\gam\dgamma}\sigma^d_{\delta\ddelta}\,R_{cdba}\label{Rcdba}\\
&&\approx 4\eps_{\gam\delta}\eps_{\beta\alpha}\Bigg[\frac 14\Big( \bar\ED_{\dgamma}\bar W_{\ddelta\dbeta\dal}
+\bar \ED_{\ddelta}\bar W_{\dbeta\dal\dgamma}+\bar\ED_{\dbeta}\bar W_{\dal\dgamma\ddelta}+
\bar\ED_{\dal}\bar W_{\dgamma\ddelta\dbeta}\Big)\0\\
&&-\frac 18 \sum_{\dgamma\ddelta}\sum_{\dbeta\dal}\eps_{\dgamma\dbeta}\sum_{\ddelta\dal}\Big[\bar\ED_{\ddelta}
\ED^{\eps} G_{\eps\dal}+\frac 14 \bar \ED_{\ddelta}\ED^{\eps}\left(\frac 13 c_{\eps\dal}-i\,n\, d_{\eps\dal}\right)+\frac i2(n-1) \bar\ED_{\ddelta} \ED_{\eps\dal}T^{\eps}\0\\
&&+n\,\ED_{\eps\ddelta} d^{\eps}{}_{\dal}+\frac 16\ED_{\eps\ddelta}c^{\eps}{}_{\dal}
 \Big]
+ (\eps_{\dgamma\dal}\eps_{\dbeta\ddelta}+\eps_{\ddelta\dal}\eps_{\dbeta\dgamma})
\Lambda\Bigg]\0\\
&&-\frac 12 \eps_{\gam\delta}\,\eps_{\dbeta\dal}\sum_{\al\beta}\sum_{\dgamma\ddelta}\Big[i\,\ED_{\beta\dgamma}\Big(G_{\al\ddelta}+\frac 13 c_{\al\ddelta}-i\,n \,d_{\al\ddelta}\Big)+\bar \ED_{\dgamma}\ED_\beta\left(G_{\al\ddelta}+ \frac 13 c_{\al\ddelta} \right)+\frac 13 \bar \ED_{\dgamma}\ED_\beta  c_{\al\ddelta} \Big]\0\\
&&-\frac 12 \eps_{\dgamma\ddelta}\,\eps_{\beta\al}\sum_{\dal\dbeta}\sum_{\gamma\delta}\Big[i\,\ED_{\gamma\dbeta}\Big(G_{\delta\dal}+\frac 13 c_{\delta\dal}-i\,n\,d_{\delta\dal} \Big)+\bar \ED_{\dbeta}\ED_\gamma\left(G_{\delta\dal}+ \frac 13 c_{\delta\dal}\right)+\frac 13 \bar \ED_{\dbeta}\ED_{\gamma}  c_{ \delta\dal} \Big]\0\\
&&+4\eps_{\dgamma\ddelta}\eps_{\dbeta\dal}\Bigg[-\frac 14\Big(\ED_{\gamma} W_{\delta\beta\al}
+ \ED_{\delta} W_{\beta\al\gamma}+\ED_{\beta} W_{\al\gamma\delta}+
\ED_{\al} W_{\gamma\delta\beta}\Big)\0\\
&&+\frac 18 \sum_{\gamma\delta}\sum_{\beta\al}\eps_{\gamma\beta}\sum_{\delta\al}\Big[\ED_{\delta}
\bar\ED^{\deps} G_{\al\deps}+\frac 14 \ED_{\delta}\ED^{\deps}\left(\frac 13 c_{\al\deps}-i\,n\, d_{\al\deps}\right)+\frac i2(n-1) \ED_{\delta} \ED_{\deps\dal}\bar T^{\deps}\0\\
&&+n\,\ED_{\delta\deps} d^{\deps}{}_{\al}+\frac 16\ED_{\delta\deps}c_{\al}^{\deps}
 \Big]
+ (\eps_{\gamma\al}\,\eps_{\beta\delta}+\eps_{\delta\al}\,\eps_{\beta\gamma})
{\boldsymbol\Lambda}\Bigg]\0
\ee
where 
\be
{\boldsymbol\Lambda} &\approx&\frac 1{24} \Big(\ED^{\al}\ED_{\al} R +\bar\ED_{\dal}\bar \ED^{\dal} R^+\Big)+\frac 1{48} \Big(\ED^{\al}\bar\ED^{\dal}-\bar\ED^{\dal}\ED^{\al}\Big) G_{\al\dal}\0\\
&& +\frac 1{24}\Big(-\frac 1{12} \ED^{\al}\bar\ED^{\dal}c_{\al\dal} -\frac i4\,n\, \ED^{\al}\bar\ED^{\dal}d_{\al\dal}-\frac i2 (n-1) \ED^{\al} \ED_{\al\dal}\bar T^{\dal} \Big) \0\\
&&+\frac 1{24}\Big(\frac 1{12} \bar\ED^{\dal}\ED^{\al} c_{\al\dal} -\frac i4\,n\, \bar\ED^{\dal}\ED^\al d_{\al\dal}+\frac i2 (n-1) \bar \ED^{\dal} \ED_{\al\dal} T^\al \Big)
-\frac n{16} \ED^{\al\dal}d_{\al\dal} \label{Lambda}
\ee

Other relations come from constraints among the various superfields
\be 
\ED^{\al}\ED_{\al} R -\bar\ED_{\dal}\bar \ED^{\dal} R^+ &\approx& -2i\, \ED^{\al\dal}G_{\al\dal}\label{DDR}\\
&&+\Big(\frac 1{12} \ED^{\al}\bar\ED^{\dal}c_{\al\dal} +\frac i4\,n\, \ED^{\al}\bar\ED^{\dal}d_{\al\dal}+\frac i2 (n-1) \ED^{\al} \ED_{\al\dal}\bar T^{\dal} \Big)\0\\
&&+\Big(\frac 1{12} \bar\ED^{\dal}\ED^{\al} c_{\al\dal} -\frac i4\,n\, \bar\ED^{\dal}\ED^\al d_{\al\dal}+\frac i2 (n-1) \bar \ED^{\dal} \ED_{\al\dal} T^\al \Big)+ i \ED^{\al\dal}c_{\al\dal}\0
\ee
together with
\be 
\ED_\al\bar\ED_{\dal}R &\approx& \frac i8 \,(n+1)\,\bar \ED_{\dal}\bar \ED_{\dbeta} d_{\al}{}^{\dbeta}-
\frac i4\,(n-1)\, \bar \ED_{\dal} \ED_{\al\dbeta}\bar T^{\dbeta}\label{DDbarR}\\
\bar\ED_{\dal}\ED_{\al}R&\approx& \bar\ED_{\dal}\bar\ED^{\dbeta}G_{\al\dbeta}-\frac i8\, n\, \bar \ED_{\dal}\bar \ED^{\dbeta} d_{\al\dbeta}+\frac 5{24}\, \bar \ED_{\dal}\bar \ED^{\dbeta} c_{\al\dbeta}-
\frac i4\,(n-1) \,\bar \ED_{\dal} \ED_{\al\dbeta}\bar T^{\dbeta}\label{DbarDR}\\
\ED^\gamma W_{\al\beta\gamma} &\approx& \frac 1{16}\left(\ED_\al\bar \ED^{\dgamma} G_{\beta\dgamma} + \ED_\beta\bar \ED^{\dgamma} G_{\al\dgamma}\right)- \frac 7{144}
\left( \ED_\al\bar \ED^{\dgamma} c_{\beta\dgamma} + \ED_\beta\bar \ED^{\dgamma} c_{\al\dgamma}\right)\label{DWab}\\
&&+\frac{in}{48} \left( \ED_\al\bar \ED^{\dgamma} d_{\beta\dgamma} + \ED_\beta\bar \ED^{\dgamma} d_{\al\dgamma}\right)\0
\ee
with the respective conjugate relations.

The last equation above, together with
\be
\ED_{\al} W_{\beta\gamma\delta}= \ED_{(\al} W_{\beta \gamma\delta)} +\frac 14 (\epsilon_{\al\beta}
\ED^\zeta W_{\gamma \delta\zeta}+\epsilon_{\al\gamma}\ED^\zeta W_{ \delta\beta\zeta}+\epsilon_{\al\delta}\ED^\zeta W_{\beta\gamma\zeta}),\label{DaWbcd1}
\ee 
allows us to conclude that
\be
\ED_{\al} W_{\beta\gamma\delta}\approx\ED_{(\al} W_{\beta \gamma\delta)}\label{DaWbcd2}
\ee

Finally we quote in its exact form a constraint equation  
\be 
\bar\ED^{\al} G_{\al\dal} -\ED_{\al}R&=& \frac 14(n+\frac 13) \bar\ED\bar\ED T_\al -\frac 14(n-\frac 13)\bar\ED^{\dal}\ED_\al \bar T_{\dal}-\frac 13 (n+1)(3n-1) \bar T^{\dal}\ED_{\dal}T_\al \0\\
&&-\frac 14(3n-5) \bar T^{\dal}G_{\al\dal} -\frac 16(6n^2+3n+1)\bar T^{\dal} \ED_\al \bar T_{\dal}+\frac 16 (n-1) T_\al \bar\ED \bar T\0\\
&&-\frac 13 (n-1) T_\al \bar T \bar T-
\frac i2 (n+1)\ED_{\al\dal}\bar T^{\dal}\label{DG}
\ee
which, together with its conjugate, is needed for the cohomological analysis of cocycles.


\end{document}